\documentclass[12pt]{article}

\usepackage{authblk} 

\usepackage{xurl} 
\usepackage{breakurl}

\usepackage{lipsum} 
\usepackage{hanging} 
\usepackage[hidelinks]{hyperref}

\title{International Agreements on AI Safety: Review and Recommendations for a Conditional AI Safety Treaty}
\author[1]{Rebecca Scholefield}
\author[2]{Samuel Martin}
\author[1]{Otto Barten}

\affil[1]{Existential Risk Observatory}
\affil[2]{Independent}

\date{} 

\begin{document}

\maketitle
\thispagestyle{empty}

\newpage
\begin{abstract}

The malicious use or malfunction of advanced general-purpose AI (GPAI) poses risks that, according to leading experts, could lead to the `marginalisation or extinction of humanity.'\footnote{Yoshua Bengio and others, \textit{International AI Safety Report} (2025) \url{https://www.gov.uk/government/publications/international-ai-safety-report-2025}, p.101.} To address these risks, there are an increasing number of proposals for international agreements on AI safety. In this paper, we review recent (2023-) proposals, identifying areas of consensus and disagreement, and drawing on related literature to assess their feasibility.\footnote{For a review of proposed  institutions, see Matthijs M. Maas and José Jaime Villalobos, ‘International AI Institutions: A Literature Review of Models, Examples, and Proposals,’ \textit{AI Foundations Report} 1 (2023) \url{http://dx.doi.org/10.2139/ssrn.4579773}.} We focus our discussion on risk thresholds, regulations, types of international agreement and five related processes: building scientific consensus, standardisation, auditing, verification and incentivisation. 

Based on this review, we propose a treaty establishing a compute threshold above which development requires rigorous oversight. This treaty would mandate complementary audits of models, information security and governance practices, overseen by an international network of AI Safety Institutes (AISIs) with authority to pause development if risks are unacceptable. Our approach combines immediately implementable measures with a flexible structure that can adapt to ongoing research.

\end{abstract}

\newpage
\section*{Acknowledgements}

We would like to thank Tolga Bilge, Oliver Guest, Jack Kelly, David Krueger, Matthijs Maas and José Jaime Villalobos for their insights.

\tableofcontents

\newpage
\section*{Introduction}

The launch of ChatGPT in 2022 amplified awareness of the potential and risks of advanced general-purpose AI (GPAI), prompting an unprecedented wave of policy responses. These include the Biden Administration's Executive Orders,\footnote{Executive Office of the President [Joe Biden], \textit{Safe, Secure, and Trustworthy Development and use of Artificial Intelligence} (2023) \url{https://www.govinfo.gov/content/pkg/FR-2023-11-01/pdf/2023-24283.pdf}.} the incorporation of provisions on GPAI into the EU AI Act, and summits held in the UK, South Korea and France. AI Safety Institutes (AISIs) or similar institutes have been established in the UK, US, EU, Japan, Singapore, France, and Canada. The UK AISI published the \textit{International AI Safety Report}, with contributions from 33 countries, in January 2025.\footnote{Bengio and others, \textit{International AI Safety Report}.}

In addition to these government-led initiatives, other international efforts have emerged, including the OECD AI Principles (2019),\footnote{OECD, \textit{AI Principles}, \url{https://www.oecd.org/en/topics/ai-principles.html} [accessed Feb 25, 2025].} the Global Partnership on AI (2020),\footnote{OECD, \textit{Global Partnership on Artificial Intelligence}, \url{https://www.oecd.org/en/about/programmes/global-partnership-on-artificial-intelligence.html} [accessed Feb 25, 2025].} UNESCO's \emph{Recommendation on the Ethics of Artificial Intelligence} (2021),\footnote{UNESCO, \textit{Recommendation on the Ethics of Artificial Intelligence}, (2022) \url{https://unesdoc.unesco.org/ark:/48223/pf0000381137}.} the G7 Hiroshima AI Process (2023),\footnote{\textit{Hiroshima Process International Code of Conduct for Advanced AI Systems} (2023) \url{https://digital-strategy.ec.europa.eu/en/library/hiroshima-process-international-code-conduct-advanced-ai-systems}; \textit{Hiroshima Process International Guiding Principles for Advanced AI Systems} (2023) \url{https://digital-strategy.ec.europa.eu/en/library/hiroshima-process-international-guiding-principles-advanced-ai-system}.} the UN High-Level Advisory Body on AI (2023), and the Council of Europe’s Framework Convention on Artificial Intelligence (2024).\footnote{Council of Europe: Committee of Ministers, \textit{Council of Europe Framework Convention on Artificial Intelligence and Human Rights, Democracy and the Rule of Law} (2024) \url{https://www.refworld.org/legal/agreements/coeministers/2024/en/148016}.}

However, many researchers argue that these efforts are insufficient. They note that, while the development of advanced AI is widely regarded as high-risk, it does not face the same level of oversight as other high-risk sectors.\footnote{Future of Life Institute, \textit{Pause Giant AI Experiments: An Open Letter} (2023) \url{https://futureoflife.org/open-letter/pause-giant-ai-experiments/} [accessed Feb 25, 2025]; Center for AI Safety, \textit{Statement on AI Risk} (2023), \url{https://www.safe.ai/work/statement-on-ai-risk} [accessed Feb 25, 2025]. See also OpenAI, \textit{OpenAI O1 System Card} (2024) \url{https://cdn.openai.com/o1-system-card-20241205.pdf}.} Rather, many initiatives rely on voluntary commitments or broadly-defined principles. There have therefore been several proposals for international agreements to mitigate the risks and maximise the benefits of AI, through concrete measures to coordinate safety, as well as research, development and benefit-sharing.\footnote{Jide Alaga and Jonas Schuett, ‘Coordinated Pausing: An Evaluation-Based Coordination Scheme for Frontier AI Developers’, (2023) \url{https://doi.org/10.48550/arXiv.2310.00374}; Duncan Cass-Beggs and others, ‘Framework Convention on Global AI Challenges’, (2024) \url{https://www.cigionline.org/publications/framework-convention-on-global-ai-challenges/}; Jason Hausenloy, Andrea Miotti and Claire Dennis, ‘Multinational AGI Consortium (MAGIC): A Proposal for International Coordination on AI’, (2023) \url{https://doi.org/10.48550/arXiv.2310.09217}; Emma Klein and Stewart Patrick, ‘Envisioning a Global Regime Complex to Govern Artificial Intelligence’, (2024) \url{https://carnegieendowment.org/research/2024/03/envisioning-a-global-regime-complex-to-govern-artificial-intelligence?lang=en}; Andrea Miotti and Akash Wasil, ‘An International Treaty to Implement a Global Compute Cap for Advanced Artificial Intelligence’, (2023a) \url{https://doi.org/10.48550/arXiv.2311.10748}; Andrea Miotti and Akash Wasil, ‘Taking Control: Policies to Address Extinction Risks from Advanced AI’, (2023b) \url{https://doi.org/10.48550/arXiv.2310.20563}; PauseAI, \textit{PauseAI Proposal} (2024) \url{https://pauseai.info/proposal}; Huw Roberts and others, ‘Global AI Governance: Barriers and Pathways Forward’, \textit{International Affairs} 100 (2024), 1275–1286 \url{https://doi.org/10.1093/ia/iiae073}; Robert Trager and others, ‘International Governance of Civilian AI: A Jurisdictional Certification Approach’, (2023) \url{https://doi.org/10.48550/arXiv.2308.15514}; \textit{Treaty on Artificial Intelligence Safety and Cooperation (TAISC)}, \url{https://taisc.org}; José Jaime Villalobos and Matthijs M. Maas, ‘Beyond a Piecemeal Approach: Prospects for a Framework Convention on AI’, \textit{SSRN}. Forthcoming in \textit{The Oxford Handbook on the Foundations and Regulation of Generative AI}, ed. by P Hacker, A Engel, S Hammer and B Mittelstadt, (Oxford: Oxford University Press, 2024), \url{https://papers.ssrn.com/abstract=5020616}.}

This paper focuses on proposals to mitigate the risks of advanced GPAI. These include risks associated with the malicious use of AI to generate fake content, commit cyber attacks, or create chemical, biological, radiological or nuclear (CBRN) weapons. They also include risks of malfunction, which encompass reliability and bias issues, and, importantly, scenarios where humans lose the ability to control AI. While experts disagree on the likelihood or imminence of these scenarios, some believe that they could result in ‘the marginalization or extinction of humanity.’\footnote{Bengio and others, \textit{International AI Safety Report}, p.101. The report lists the following capabilities associated with loss of control: agent capabilities, scheming, theory of mind, situational awareness, persuasion, autonomous replication and adaptation, AI development, offensive cyber capabilities, and general R\&D.}

Researchers recommend international agreements over domestic regulation for several reasons.\footnote{See Trager and others, ‘International Governance of Civilian AI,’ p.2} Firstly, verifying compliance with agreements will likely require international cooperation, as the AI hardware supply chain is global. Secondly, since AI-related harms cannot be contained within jurisdictional boundaries, there is a strong case for international collaboration. Thirdly, without this collaboration, competitive pressures between states could lead to the neglect of safety.

We begin this paper by reviewing proposals for international agreements. We identify their common components and draw on related literature for additional context and insights. In Section 1.1 and Section 1.2, we discuss proposed risk thresholds and the safety assurances researchers believe should be required above these thresholds. We then focus on issues related to implementation. In Section 1.3, we discuss legal and alternative approaches to increasing international cooperation, and in Section 1.4, we look at five relevant processes: building scientific consensus, standardisation, auditing, verification and incentivisation.

In Section 2, we draw from the review to discuss aspects of international agreements that can be readily implemented and those that require further research. We then offer recommendations. 

We exclude systemic risks—such as risks to the labour market and environment, and global inequality—from the scope of this paper, as the proposals we discuss are not directly applicable to them.\footnote{See Bengio and others, International AI Safety Report, p.110-148.}  We nonetheless acknowledge their importance.

\addcontentsline{toc}{section}{Introduction}
\section{Review}

\subsection{Risk Thresholds}

Most proposals aim to regulate the development of AI models that use more than a certain amount of computation during training. Computation, or compute, can be measured in floating-point operations (FLOP) and calculated based on the `number of computational operations used in training.'\footnote{In practice, there is disagreement about whether compute thresholds should be calculated based solely on pre-training compute, or whether they should also account for compute used in post-training enhancement and deployment/inference. See Sara Hooker, ‘On the Limitations of Compute Thresholds as a Governance Strategy,’ (2024) \url{https://doi.org/10.48550/arXiv.2407.05694}, p.13. Cf. Lennart Heim and Leonie Koessler, ‘Training Compute Thresholds: Features and Functions in AI Regulation’ (2024) \url{https://doi.org/10.48550/arXiv.2405.10799}, p.7; Lennart Heim, \textit{Inference Compute: GPT-o1 and AI Governance}, (2024) \url{https://blog.heim.xyz/inference-compute/} [accessed Feb 25, 2025].} In other words, proposals set a training compute threshold and regulate AI that exceeds it.

Training compute thresholds can act as an `initial filter' for identifying models that may pose a high risk.\footnote{Heim and Koessler, ‘Training Compute Thresholds,’ p.3.} This is because increases in training compute tend to correlate with increases in a model’s capabilities, a phenomenon attributed to empirical `scaling laws.'\footnote{Ibid., p.2. See also Bengio and others, \textit{International AI Safety Report}, p.51-52.}  Increased capabilities can, in turn, be associated with increased risk.

Researchers also point to the practical advantages of training compute thresholds. Compute, as a physical resource in a concentrated supply chain, is `\emph{detectable, excludable}, and \emph{quantifiable}.'\footnote{Girish Sastry and others, ‘Computing Power and the Governance of Artificial Intelligence’, (2024) \url{https://doi.org/10.48550/arXiv.2402.08797}, p.1}

Training compute thresholds already appear in high-profile proposals and legislation, setting a precedent for proposals for international agreements. For example, the EU Act defines `general purpose AI models with systemic risk' as those trained with more than 10\textsuperscript{25} FLOP.\footnote{‘Regulation (EU) 2024/1689 of the European Parliament and of the Council of 13 June 2024 Laying Down Harmonised Rules on Artificial Intelligence and Amending Regulations,’ \textit{Official Journal} L (2024) \url{https://eur-lex.europa.eu/legal-content/EN/TXT/PDF/?uri=OJ:L_202401689}, p. 83.} The Biden Administration’s Executive Order (EO) 14110 imposed reporting requirements on developers training models with more than 10\textsuperscript{26} FLOP.\footnote{Executive Office of the President, \textit{Safe, Secure, and Trustworthy Development and use of Artificial Intelligence}, p.75197.} The vetoed Senate Bill (SB) 1047 would have initially covered models trained with more than 10\textsuperscript{26} FLOP.\footnote{Scott Wiener, \textit{Safe and Secure Innovation for Frontier Artificial Intelligence Models Act}, (2024)\url{https://leginfo.legislature.ca.gov/faces/billTextClient.xhtml?bill_id=202320240SB1047}}. 

However, proposals for international agreements generally advocate for lower thresholds. Miotti and Wasil propose 10\textsuperscript{21} FLOP;\footnote{Miotti and Wasil, ‘An International Treaty to Implement a Global Compute Cap for Advanced Artificial Intelligence’, p.7.} Bilge proposes 10\textsuperscript{23} FLOP;\footnote{\textit{Treaty on Artificial Intelligence Safety and Cooperation.}} Trager and others suggest 10\textsuperscript{24} FLOP;\footnote{Trager and others, ‘International Governance of Civilian AI,’ p.27.} and PauseAI proposes 10\textsuperscript{25} FLOP.\footnote{PauseAI, \textit{PauseAI Proposal}.}

Proposals also note that thresholds would need to be periodically revised, as improvements to algorithmic efficiency would reduce the amount of compute required to train a given model. For example, in their ‘Treaty on the Prohibition of Dangerous Artificial Intelligence,’ Miotti and Wasil propose that State Parties meet at least once a year in Geneva to review the threshold.\footnote{Miotti and Wasil, ‘An International Treaty to Implement a Global Compute Cap for Advanced Artificial Intelligence’, p.7.} Heim and Koessler argue thresholds should be revised as often as “every couple of months.\footnote{Heim and Koessler, ‘Training Compute Thresholds,’ p.22.} There are also precedents for revising thresholds in the EU AI Act, EO 14110 and SB 1047, which delegate the process to the EU AI Office, US Secretary of Commerce and California Government Operations Agency, respectively.\footnote{‘Regulation (EU) 2024/1689,’ p.29; Executive Office of the President, \textit{Safe, Secure, and Trustworthy Development and use of Artificial Intelligence}, p.75197; Wiener, \textit{Safe and Secure Innovation for Frontier Artificial Intelligence Models Act}.}

While most proposals use training compute thresholds, some researchers emphasise that scaling laws are based on empirical observation rather than scientific laws, which limits their effectiveness as a proxy for risk.\footnote{Hooker, ‘On the Limitations of Compute Thresholds as a Governance Strategy,’ p.18-21. See also Bengio and others, \textit{International AI Safety Report}, p.52-3.} Consequently, they advocate for thresholds that incorporate multiple metrics, such as:
\begin{itemize}
\item \textbf{Model architecture and training algorithms}: Heim and Koessler note that these are `hard to quantify.'\footnote{Heim and Koessler, ‘Training Compute Thresholds,’ p.26.}
\item \textbf{Number of model parameters}: Heim and Koessler, and Miotti and others, note that the number of parameters correlates with training compute, and argue training compute is a preferable metric as it is more monitorable. However, PauseAI includes parameters in its threshold.
\item \textbf{Amount and quality of training data}: Heim and Koessler note there are no `objective or standardized' methods to measure these metrics. However, Miotti and others seem to see them as more promising and mention future work to discuss options.\footnote{Andrea Miotti, Tolga Bilge, Dave Kasten and James Newport, A Narrow Path: \textit{How to secure our future} (2024) \url{https://pdf.narrowpath.co/A_Narrow_Path.pdf}, p.78.}
\item \textbf{Estimated capabilities}: PauseAI includes `capabilities that are expected to exceed GPT-4' in their threshold, but note that capabilities are hard to predict and that this threshold may be hard to enforce.\footnote{PauseAI, \textit{PauseAI Proposal}.}
\end{itemize}

Some proposals stop at one threshold and one set of requirements. Others propose a second, higher threshold, above which development would only be permitted in a centralised, international institution. For example, Bilge proposes a `Joint AI Safety Laboratory' (JAISL) to develop models above a threshold of 10\textsuperscript{23} FLOP.\footnote{\textit{Treaty on Artificial Intelligence Safety and Cooperation}.} Miotti and Wasil propose a `Multinational AGI Consortium' (MAGIC) to develop models above a threshold of 10\textsuperscript{24} FLOP.\footnote{Miotti and Wasil, ‘Taking control,’ p.7. See also Hausenloy, Miotti and Dennis, ‘Multinational AGI Consortium (MAGIC).’} Cass-Beggs and others propose an `international joint AI lab' to develop models above a qualitative `tolerable' risk threshold.\footnote{Cass-Beggs and others, ‘Framework Convention on Global AI Challenges,’ p.15.} In addition to safely developing advanced AI, Bilge notes that the JAISL would research AI safety and `the alignment problem.'\footnote{\textit{Treaty on Artificial Intelligence Safety and Cooperation}.}

Some proposals include a third threshold above which development would be unconditionally prohibited. Bilge proposes a threshold of 2.5 x 10\textsuperscript{25} FLOP, and Cass Beggs and others propose an `unacceptable risk' threshold, above which development would be prohibited “until adequate safety and control mechanisms become available.\footnote{Ibid.; Cass-Beggs and others, ‘Framework Convention on Global AI Challenges,’ p.16.}

\subsection{Regulation}

As stated above, thresholds—whether based on training compute or other metrics—are used as an `initial filter' to identify models that should be regulated.\footnote{Heim and Koessler, ‘Training Compute Thresholds,’ p.3.} Proposed regulation often involves requirements for models and an organisation’s security and governance practices. 

\subsubsection{Models}

In many proposals, evaluation results determine whether a model may be further developed or deployed.\footnote{Alaga and Schuett, ‘Coordinated Pausing;’ Cass Beggs and others, ‘Framework Convention on Global AI Challenges,’ p.15; PauseAI, \textit{PauseAI Proposal}.} Alaga and Schuett and PauseAI refer specifically to evaluating \emph{dangerous} capabilities, which may include ‘cyber-offense, deception, persuasion \& manipulation, political strategy, weapons acquisition, long-horizon planning, AI development, situational awareness, and self-proliferation.’\footnote{Toby Shevlane and others, ‘Model Evaluation for Extreme Risks’, (2023) \url{https://doi.org/10.48550/arXiv.2305.15324}, p.5.} Anthropic, Google DeepMind, Meta and OpenAI report evaluating some of these capabilities in their model and system cards, and refer to them in risk management policies.\footnote{See Anthropic, \textit{Anthropic's Responsible Scaling Policy}, (2023) \url{https://www-cdn.anthropic.com/files/4zrzovbb/website/1adf000c8f675958c2ee23805d91aaade1cd4613.pdf}; OpenAI, \textit{Preparedness Framework (Beta)}, (2023) \url{https://cdn.openai.com/openai-preparedness-framework-beta.pdf}}

However, evaluations are generally considered an immature science—or even an `art'.\footnote{Appollo Research, \textit{We need a Science of Evals} (2024) \url{https://www.apolloresearch.ai/blog/we-need-a-science-of-evals} [accessed Feb 25, 2025].} It is not currently known how to create evaluations that are valid, reliable and comprehensive.\footnote{Bengio et al, \textit{International AI Safety Report}, p.184.} For example, changes to prompts used in evaluations can significantly impact how systems perform against benchmarks. It is also impossible to anticipate all the scenarios in which AI could cause harm, and all the dangerous capabilities that could enable those scenarios.\footnote{Miotti and Wasil, ‘Taking Control,’ p.6.} Therefore, evaluations cannot provide total confidence in a model’s safety—nor do researchers claim they can. As the UK AISI has stated, evaluations cannot currently ‘act as a "certification" function (i.e., provide confident assurances that a particular system is "safe").’\footnote{UK AI Security Institute, \textit{Early Lessons from Evaluating Frontier AI Systems}, (2024) \url{https://www.aisi.gov.uk/work/early-lessons-from-evaluating-frontier-ai-systems} [accessed Feb 25, 2025].}

Nevertheless, many organisations and AISIs are working to advance the science of evaluations. This reflects the important role that they play in existing approaches to regulation.\footnote{Anka Reuel and others, ‘Position Paper: Technical Research and Talent is Needed for Effective AI Governance’, \textit{Proceedings of the 41st International Conference on Machine Learning} (June 11, 2024) \url{https://doi.org/10.48550/arXiv.2406.06987}, p.5.}

A more fundamental critique of evaluations is that it may be unclear, for policymaking purposes, what a model’s performance signifies. According to Righetti, a model that `fails' an evaluation for a dangerous capability is likely to be safe (in relation to the associated risk), but a model that `passes' the same evaluation is not necessarily dangerous.\footnote{Luca Righetti, \textit{Dangerous capability tests should be harder} (2024) \url{https://www.planned-obsolescence.org/dangerous-capability-tests-should-be-harder/} [accessed Feb 25, 2025].} For example, a model that can answer questions about biology cannot necessarily help a non-expert create a bioweapon. Taking a more pragmatic approach, Righetti recommends developing tests that would unambiguously persuade policymakers and the public that an AI poses enough risk to justify interventions like pausing development. He offers the hypothetical example of a randomised controlled trial to test whether non-experts could create bioweapons. He suggests `thinking backwards' from such examples to create tests that are both practical and convincing.

In contrast to Righetti, who assumes that policymakers will need to prove models are dangerous, Miotti and Wasil argue that the onus should be on developers to provide `affirmative evidence of safety.'\footnote{Miotti and Wasil, ‘Taking Control,’ p.5.} They emphasise that this is a common practice in high-risk sectors. Developers would ideally be required to prove they understand, for example, how a system `reaches conclusions.' However, Miotti and Wasil acknowledge that this is not possible in practice. They suggest that, in reality, broader risk assessments may be used to determine whether risks are kept beneath certain levels. Similarly, Wasil and others provide examples of affirmative evidence about model outputs, internals and training processes, but acknowledge that relevant research is in early stages.\footnote{Akash Wasil and others, ‘Affirmative Safety: An Approach to Risk Management for High-Risk AI’, (2024) \url{https://doi.org/10.48550/arXiv.2406.15371}, p.7.}

\subsubsection{Security}

Audits of information security (including cybersecurity) and physical security take place in many high-risk industries. For example, in the USA, nuclear power plants are inspected to ensure compliance with cybersecurity regulations.\footnote{United States Nuclear Regulatory Commission, \textit{Cybersecurity} (2025) \url{https://www.nrc.gov/security/cybersecurity.html} [accessed 25 Feb, 2025].}

Many proposals recommend information security requirements for companies developing AI models above a certain threshold.\footnote{Cass-Beggs and others, ‘Framework Convention on Global AI Challenges,’ p.15; Miotti and Wasil, ‘An International Treaty to Implement a Global Compute Cap for Advanced Artificial Intelligence,’ p.7; Trager and others, ‘International Governance of Civilian AI,’ p.27-28.} This is due, for example, to the risk that cyberattackers access model components such as source code, model weights and training data. This could enable them to increase a model’s dangerous capabilities by removing safety filters, and fine-tuning and jailbreaking models.\footnote{Elizabeth Seger and others, ‘Open-Sourcing Highly Capable Foundation Models: An Evaluation of Risks, Benefits, and Alternative Methods for Pursuing Open-Source Objectives’, (2023) \url{https://doi.org/10.48550/arXiv.2311.09227}, p.12.}

\subsubsection{Governance}

Auditors can also assess whether organisations meet `procedural prescriptions' related to safety—for example, by reviewing risk assessment, risk mitigation, and emergency response procedures—or have a strong \textit{safety culture}.\footnote{Merlin Stein and others, ‘Public Vs Private Bodies: Who should Run Advanced AI Evaluations and Audits? A Three-Step Logic Based on Case Studies of High-Risk Industries,’ (2024) \url{https://www.oxfordmartin.ox.ac.uk/publications/public-vs-private-bodies-who-should-run-advanced-ai-evaluations-and-audits-a-three-step-logic-based-on-case-studies-of-high-risk-industries}, p.11.}

The concept of \textit{safety culture} was developed in the nuclear industry, and is defined by the International Atomic Energy as an:

\begin{quote}
assembly of characteristics, attitudes and behaviours in individuals, organizations and institutions which establishes that, as an overriding priority, protection and safety issues receive the attention warranted by their significance.\footnote{International Atomic Energy Agency, \textit{Safety and Security Culture} (2016) \url{https://www.iaea.org/topics/safety-and-security-culture} [accessed Feb 25, 2025]; David Manheim, `Building a Culture of Safety for AI: Perspectives and Challenges', (2023) \url{https://papers.ssrn.com/abstract=4491421} [accessed Feb 25, 2025], p.2.}
\end{quote}

To assess the safety culture at nuclear facilities, the International Atomic Energy Agency performs on-site visits to review documents and procedures and interview staff. It aims to investigate employees’ attitudes towards safety and understanding of risks, as well as communication and resource allocation within an organisation.
Researchers argue that, since advanced AI development is considered high-risk, audits of developers’ risk management procedures and safety culture should be common practice.

\subsubsection{Arguments}

A combination of evidence related to models, security and governance could be considered when deciding whether a model may be developed above a certain threshold. 

Researchers have also discussed the possibility of integrating this evidence into a safety case, defined as a:

\begin{quote}
Structured argument supported by evidence, where the developer identifies hazards, models risk scenarios, and evaluates the mitigations taken [...] To demonstrate that their product does not exceed maximum risk thresholds set by the regulator.\footnote{Bengio and others, \textit{International AI Safety Report}, p.167.}
\end{quote}

By focusing on risk thresholds, safety cases can accommodate a range of evidence. For example, Buhl and others suggest that, in the short term, safety cases may use evaluations to evidence that a system lacks a certain capability. In the longer term, however, evidence may consist of `\emph{mathematical models}' and `\emph{formal verifications or proofs}.'\footnote{Marie Davidsen Buhl and others, ‘Safety cases for frontier AI’ (2024) \url{<https://doi.org/10.48550/arXiv.2410.21572}.}

The UK AISI has stated an interest in safety cases and is working with `two frontier AI labs,' Apollo Research, Redwood Research and the Centre for the Governance of AI (GovAI) to create provisional safety case `sketches.'\footnote{Geoffrey Irving, \textit{Safety Cases at AISI}, (2024) \url{https://www.aisi.gov.uk/work/safety-cases-at-aisi} [accessed Feb 25, 2025].} So far, the AISI has published a sketch for `offensive cyber capabilities' in collaboration with GovAI.\footnote{Arthus Goemans and others, `Safety Case Template for Frontier AI: A Cyber Inability Argument', (2024) \url{https://doi.org/10.48550/arXiv.2411.08088}.}

\subsection{Types of Agreement}

Those who propose thresholds and regulations mainly propose enforcing these via treaties, a source of international law. Others argue that ratifying and implementing treaties will be infeasible in the short to medium term. Instead, they recommend coordinating efforts within the existing `regime complex' of AI governance initiatives and institutions. To this end, they advocate for `soft law,' which includes voluntary resolutions, recommendations, codes of conduct, and standards.\footnote{Teresa Fajardo, ‘Soft Law,’ \textit{Oxford Bibliographies}, (2014) \url{https://www.oxfordbibliographies.com/display/document/obo-9780199796953/obo-9780199796953-0040.xml} [accessed Feb 25, 2025].}

\subsubsection{Treaties}

Some proposals are written in the style of multilateral treaties that establish international institutions. Multilateral treaties are those with more than two states parties. 

For example, PauseAI proposes a treaty to establish an ‘international AI safety agency;’\footnote{PauseAI, \textit{PauseAI Proposal}.} Tolga Bilge proposes a ‘Treaty on Artificial Intelligence Safety and Cooperation’ to establish an ‘International AI Safety and Cooperation Commission;’\footnote{\textit{Treaty on Artificial Intelligence Safety and Cooperation (TAISC)}.} and Miotti and Wasil propose a UN ‘Treaty on the Prohibition of Dangerous Artificial Intelligence’ to establish ‘an international organization for monitoring, enforcement, and research.’\footnote{Miotti and Wasil, ‘An International Treaty to Implement a Global Compute Cap for Advanced Artificial Intelligence.’}

A relevant example of an existing institution is the International Atomic Energy Agency (IAEA), an autonomous organisation within the United Nations system. The IAEA was established by a statute approved at a UN conference in 1956. Its mandate was later expanded under the Treaty on the Non-Proliferation of Nuclear Weapons in 1968.

Although multilateral treaties are often seen as an ideal solution, some proposals suggest that bilateral treaties (treaties with two states parties) may be more feasible in the shorter term. For example, PauseAI states that the involvement of the US and China is a priority, followed by the EU.

\subsubsection{Framework Conventions}

Maas and Villalobos caution that ratifying multiple treaties for specific issues within AI governance could result in a `fragmented international legal regime.'\footnote{Villalobos and Maas, ‘Beyond a Piecemeal Approach,’ p.13.} Instead, they recommend framework conventions, a type of treaty. Framework conventions do not have a formal definition, but they are generally multilateral, broad in scope, and establish:

\begin{quote}
A two-step regulatory process through which initially underspecified obligations and implementation mechanisms are subsequently specified via protocols.\footnote{Ibid., p.10; Nele Matz-Lück, ‘Framework Conventions as Regulatory Tools’, \textit{Goettingen Journal of International Law}, 1 (2009), 439–458 \url{https://papers.ssrn.com/abstract=1535892}, p.441.}
\end{quote}

A well-known example is the UN Framework Convention on Climate Change (UNFCCC) (1994). Its objective is the `stabilization of greenhouse gas concentrations in the atmosphere.'\footnote{UN General Assembly, \textit{United Nations Framework Convention on Climate Change : resolution / adopted by the General Assembly}, (1994) \url{https://unfccc.int/resource/docs/convkp/conveng.pdf}, p.4.} The UNFCCC outlines how this objective can be achieved with principles, commitments, and articles that expand on certain commitments. It also establishes a Conference of Parties to adopt protocols. Another example is the Council of Europe’s Framework Convention on Artificial Intelligence and Human Rights, Democracy and the Rule of Law (2024), the first legally binding international treaty on AI.\footnote{Council of Europe: Committee of Ministers, \textit{Council of Europe Framework Convention on Artificial Intelligence and Human Rights, Democracy and the Rule of Law}.}

Maas and Villalobos highlight several advantages of framework conventions. Their broad scope enables them to coordinate multiple issues, including safety, research, standardisation and benefit sharing.\footnote{Villalobos and Maas, ‘Beyond a Piecemeal Approach,’ p.15.} Framework conventions can also establish obligations with varying levels of specificity, to account for varying degrees of `political will' and `technical certainty.'\footnote{Ibid., p.16.}
Acknowledging the risk that a framework convention on AI may be \emph{too} broad, Maas and Villalobos recommend listing potential protocols upfront, with timelines for negotiating them.\footnote{Ibid.} They also advise specifying verification and implementation mechanisms upfront and establishing an international institution to perform these functions. Finally, to encourage participation, they suggest economic incentives like market access and benefit-sharing.\footnote{Ibid., p.17.}

Cass-Beggs and others also propose a ‘Framework Convention on Global AI Challenges.’ They list nine potential protocols to achieve the `sub-objectives' of `realizing and sharing global benefits',  `addressing global AI risks' and `making globally legitimate and effective decisions about how to govern advanced AI.'\footnote{Cass-Beggs and others, ‘Framework Convention on Global AI Challenges,’ p.13.}

\subsubsection{Alternatives}

Trager and others acknowledge that de facto international agreements can be reached without treaties. These agreements can involve international institutions. For example, they refer to the Internet Corporation for Assigned Names and Numbers and International Organization for Standardization, whose standards influence regulation, as potential models for an International AI Organisation (IAIO).\footnote{Trager and others, ‘International Governance of Civilian AI,’ p.29.} The proposed IAIO would certify states’ compliance with international standards.\footnote{Ibid.} To be certified, states would also have to enforce restrictions on trading AI products with uncertified states. 

Finally, some researchers argue that it will be infeasible to create either international agreements or institutions in the near future. They emphasise obstacles like `geopolitical and economic competition,' divergent policy approaches among the US, China and the EU, and the `painstaking process' of negotiating and ratifying agreements.\footnote{Klein and Patrick, ‘Envisioning a Global Regime Complex to Govern Artificial Intelligence’, p.22; p.7.}

They therefore recommend strengthening the existing `regime complex' of initiatives and institutions related to AI governance as a more realistic alternative.\footnote{Roberts and others, ‘Global AI Governance’, p.2.} This regime complex includes international organisations like the UN, UNESCO, G7, BRICS and Council of Europe; technical standards bodies; specialised initiatives like the Global Partnership on AI and AI Safety Institutes; and events like the international AI summits.

For example, Klein and Patrick envisage a `\emph{multi-multilateral}' regime complex, in which different states join different initiatives and institutions.\footnote{Klein and Patrick, ‘Envisioning a Global Regime Complex to Govern Artificial Intelligence,’ p.3.} They suggest that short-term progress will consist of `nonbinding agreements,' `declarations of principles,' and the `promotion of norms,' which could pave the way for future international agreements. Robert and others add that an expert body could unofficially lead these efforts by identifying common goals, to facilitate information exchange and institutional partnerships.\footnote{Roberts and others, ‘Global AI Governance,’ p.13.}

Both Klein and Patrick, and Roberts and others, refer to the example of climate change, which encompasses multiple policy issues, such as biodiversity and carbon emissions.\footnote{Ibid., p.11; Klein and Patrick, ‘Envisioning a Global Regime Complex to Govern Artificial Intelligence,’ p.6.} The regime complex for climate change consists of multilateral treaties, scientific assessment bodies, UN agencies, minilateral groups and private sector coalitions. 

\subsection{Related Processes}

In this section, we take a broader look at five key processes that underpin the proposals we discuss: building scientific consensus, setting standards, auditing, and verifying and incentivising compliance. We examine the literature on these processes and the roles of international institutions, governments, and private sector actors.

\subsubsection{Building Scientific Consensus}

Pouget and Dennis note that `shared scientific understanding' has been a `precondition for progress' on issues such as climate change and biodiversity.\footnote{Hadrien Pouget and Claire Dennis, ‘The Future of International Scientific Assessments of AI’s Risks’, (2024) \url{https://www.oxfordmartin.ox.ac.uk/publications/the-future-of-international-scientific-assessments-of-ais-risks}, p.5.} In the context of AI, they recommend producing multiple, complementary reports with varied contributors to account for the varying priorities and expertise of different states.

Two processes are seen as particularly promising. Firstly, the UN's Global Digital Compact commits to establishing a ‘multidisciplinary Independent International Scientific Panel on AI.’\footnote{United Nations, \textit{Global Digital Compact}, (2024) \url{https://www.un.org/global-digital-compact/en}, p.13.} Building on this commitment, Pouget and Dennis recommend that the panel reports on the risks \emph{and} benefits of AI, to engage as many member states as possible.\footnote{Pouget and Dennis, ‘The Future of International Scientific Assessments of AI’s Risks,’ p.9.} They also recommend involving policymakers, similar to the Intergovernmental Panel on Climate Change.

Secondly, many researchers recommend continuing the work of the \emph{International AI Safety Report}, commissioned by the UK government and chaired by Yoshua Bengio, with a focus on the risks of advanced AI.\footnote{Ibid., p.11} Researchers at the Oxford Martin School recommend AISIs worldwide should collaborate to produce annual or biannual reports.\footnote{Marta Ziosi and others, ‘AISIs’ Roles in Domestic and International Governance’ (2024) \url{https://www.oxfordmartin.ox.ac.uk/publications/aisis-roles-in-domestic-and-international-governance}, p.9.} AISIs, or similar institutions, exist in the US, Japan, EU, Canada, France and Singapore. China does not have an official, national AISI, although researchers have identified five Chinese institutions that, between them, fulfil similar functions.\footnote{Karson Elmgren and Oliver Guest, ‘Chinese AI Safety Institute Counterparts,’ (2024) \url{https://www.iaps.ai/research/china-aisi-counterparts}. These institutions are CAICT, Shanghai AI Lab, TC260, Institute for AI International Governance, and Beijing Academy of Artificial Intelligence.}

\subsubsection{Standardisation}

Future regulation may also reference standards with which developers must comply.\footnote{Hadrien Pouget, ‘What will the role of standards be in AI governance?’ (2023) \url{https://www.adalovelaceinstitute.org/blog/role-of-standards-in-ai-governance/}.} Researchers have discussed several approaches to developing international standards—for example, for terminology, evaluations and risk management.\footnote{National Institute of Standards and Technology, \textit{A Plan for Global Engagement on AI Standards} (2024) \url{https://doi.org/10.6028/NIST.AI.100-5}.}

Firstly, existing international standards bodies will likely continue their work to develop standards. Notably, the International Organization for Standardization (ISO) and the International Electrotechnical Commission (IEC) established a joint technical subcommittee on artificial intelligence (ISO/IEC JTC 1/SC 42). Each state can nominate one private or public body to be a member of the ISO.\footnote{International Organization for Standardization, \textit{ISO Membership Manual} (2015) \url{https://www.iso.org/publication/PUB100399.html}, p.6.} This member can, however, delegate relevant tasks to other organisations. Kristina Fort recommends that AISIs work with their states’ members to provide technical expertise as input into this standardisation process.

Secondly, Fort recommends that AISIs and similar institutions themselves develop standards. She argues that the AISIs’ expertise will enable them to work more quickly than the ISO/IEC JTC to develop standards for performance, measurement, processes and management, even if they may be perceived as a less legitimate standards body.

Thirdly, Trager and others propose an International AI Organisation (IAIO) to develop standards and monitor compliance.\footnote{Trager and others, ‘International Governance of Civilian AI.’} They note that this organisation could be part of an intergovernmental body such as the UN, or be an independent intergovernmental or non-governmental organisation.

\subsubsection{Auditing}

As noted in Section 1.2, proposals often state requirements for models, and their developers’ security and governance practices. Auditing can be used to assess whether models and developers meet these requirements.

Audits of models consist of evaluations and red teaming and can be conducted by developers, third-party private bodies or public bodies. 

Developers are the experts on their models, and some developers claim that this expertise enables them to more successfully evaluate models, for example by using prompt engineering to elicit better performance.\footnote{Anthropic, \textit{Challenges in evaluating AI systems} (2023) \url{https://www.anthropic.com/news/evaluating-ai-systems} [accessed 25 Feb, 2025].} However, researchers note that developers may face conflicts of interest in designing and executing evaluations that reveal risks presented by their products.\footnote{Ibid.; Lara Thurnherr and others, ‘Who Should Develop Which AI Evaluations,’ (2025) \url{https://www.oxfordmartin.ox.ac.uk/publications/who-should-develop-which-ai-evaluations}, p.5.} They therefore recommend third-party evaluations.

The current ecosystem of private auditors includes the evaluation and red-teaming organisations Apollo Research, Faculty, Gray Swan AI, Haize Labs, METR and VirtueAI. Researchers recommend accrediting these private auditors, for example, based on their governance practices and compliance with standards.\footnote{Ibid., p.5; Merlin Stein and others, ‘Public vs Private Bodies: Who Should Run Advanced AI Evaluations and Audits?’ (2024) \url{https://www.oxfordmartin.ox.ac.uk/publications/public-vs-private-bodies-who-should-run-advanced-ai-evaluations-and-audits-a-three-step-logic-based-on-case-studies-of-high-risk-industries}, p.13.} This would establish a competitive `regulatory market,' which Hadfield and Clark argue would encourage innovation.\footnote{Gillian Hadfield and Jack Clark, ‘Regulatory Markets: The Future of AI Governance’ (2023) \url{https://doi.org/10.48550/arXiv.2304.04914}.}

Stein and others recommend private audits where the level of risk and sensitivity of information is low and the volume of audits required is high.\footnote{Stein and others, ‘Public vs Private Bodies,’ p.11.} For example, they propose that private auditors conduct benchmark evaluations to assess models’ performance on standardised tasks. These auditors would only require black-box access, whereby they interact with a model like a normal user and assess its outputs.\footnote{Stephen Casper and Carson Ezell, ‘Black-Box Access is Insufficient for Rigorous AI Audits’ (2024) \url{https://doi.org/10.48550/arXiv.2401.14446}, p.2.}

Other evaluations require grey or white-box access, allowing auditors to examine a model's `inner workings’.\footnote{Ibid.} At most, this includes weights, activations, and gradients, and the ability to fine-tune the model.\footnote{Ibid., p.3.} Stein and others argue that these evaluations, which may concern ‘critical risks,’ should be performed by public auditors with appropriate security clearances.\footnote{Stein and others, ‘Public vs Private Bodies,’ p.13.}

AISIs currently act as public auditors. For example, the UK and US AISIs have collaborated to evaluate OpenAI’s o1 and Anthropic’s Claude 3.5 Sonnet.\footnote{\textit{US AISI and UK AISI Joint Pre-Deployment Test: Anthropic's Claude 3.5 Sonnet; US AISI and UK AISI Joint Pre-Deployment Test: OpenAI O1}.} Under an international agreement, AISIs would likely need to be able to mutually recognise each other’s evaluation results. As Ziosi and others explain, this is because AISIs may only be able to conduct evaluations requiring high levels of model access and security clearance on models developed in their jurisdiction.\footnote{Ziosi and others, ‘AISIs’ Roles in Domestic and International Governance,’ p.6.} They also recommend establishing `regional AISIs' to support states that lack AISIs or the expertise and resources necessary for auditing.

Compared to model audits, security and governance audits are more established and standardised in other high-risk industries.\footnote{Stein and others, ‘Public vs Private Bodies,’ p.13.} Stein and others suggest these audits are less likely to require access to sensitive information. They recommend that private bodies conduct them, subject to public oversight.

\subsubsection{Verification}

As mentioned in Section 1.1, compute thresholds are popular partly because compute is ‘\emph{detectable}, \emph{excludable}, and \emph{quantifiable}’ resource.\footnote{Sastry and others, ‘Computing Power and the Governance of Artificial Intelligence’, p.1.} This makes monitoring and controlling access to compute (compute governance) an effective means of verifying compliance with regulations.

One method of monitoring is to impose reporting requirements on actors in the compute supply chain. Sastry and others recommend that chip producers and sellers report transfers of AI chips, to monitor the distribution of compute among actors like semiconductor fabrication plants, assembly and test firms, and end users including cloud compute providers.\footnote{Ibid., p.39.} 

Heim and others also recommend reporting requirements for cloud compute providers, which own the infrastructure required to train frontier models.\footnote{ Lennart Heim and others, ‘Governing Through the Cloud: The Intermediary Role of Compute Providers in AI Regulation’, (2024) \url{https://www.oxfordmartin.ox.ac.uk/publications/governing-through-the-cloud-the-intermediary-role-of-compute-providers-in-ai-regulation}, p.9.} They could report the amount of compute consumed by developers using their services. This would enable regulators to identify developers training models above specified thresholds and ensure they are following regulations.

In addition to monitoring, cloud compute providers could enforce regulations by restricting access to their services. For example, Egan and Heim propose Know-Your-Customer requirements for compute providers in the US, drawing from practices in the financial sector.\footnote{Janet Egan and Lennart Heim, ‘Oversight for Frontier AI through a Know-Your-Customer Scheme for Compute Providers’, (2023) \url{https://doi.org/10.48550/arXiv.2310.13625}, p.7.} Providers would be required to report information about a company, its personnel and intended use cases, and report customers that match government-defined “high-risk” profiles.\footnote{ Ibid., p.11-12.} Similarly, Trager and others propose requiring firms to obtain licenses to train models above a certain compute threshold. Compute providers could then refuse their services to unlicensed firms.\footnote{Trager and others, ‘International Governance of Civilian AI,’ p.28.}

Researchers also propose hardware-based methods of monitoring, such as embedding `physical unique identifiers' in chips to enable location tracking, and recording them in an international registry.\footnote{Sastry and others, ‘Computing Power and the Governance of Artificial Intelligence’, p.39.} Wasil and others discuss mechanisms that could be implemented in chip firmware and drivers to detect unauthorized activities, such as unauthorized chip clustering.\footnote{Ibid.} However, further research into the technical feasibility of these solutions is required, and they may require phasing out or retrofitting hardware already in circulation.\footnote{Ibid.; Akash Wasil and others, ‘Verification methods for international AI agreements’, (2023) \url{https://doi.org/10.48550/arXiv.2408.16074}, Figure 5.}

The methods discussed above depend on compute providers cooperating with authorities. Authorities could also use surveillance methods to identify and oversee compute providers within their jurisdictions. These include satellite-based remote sensing and infrared imaging to detect data centers, as well as energy monitoring to track consumption patterns. Additionally, customs data analysis and financial intelligence could be used to trace the movement of hardware components. Finally, authorities could conduct routine or challenge inspections of compute providers’ and developers’ facilities.\footnote{Ibid.} For comparison, the  Organization for the Prohibition of Chemical Weapons can conduct challenge inspections when signatories of the Chemical Weapons Convention suspect non-compliance.

\subsubsection{Incentivisation}

While the primary motivation for an international agreement on AI would be risk mitigation, researchers have proposed additional incentives to attract signatories.

Firstly, an international research initiative could be established to pool resources and talent to research and develop AI. As mentioned in Section 1.1, proposals refer to a `Joint AI Safety Laboratory' (JAISL), `Multinational AGI Consortium' (MAGIC) and `international joint AI lab'.\footnote{\textit{Treaty on Artificial Intelligence Safety and Cooperation (TAISC)}; Miotti and Wasil, ‘Taking control,’ p.7; Hausenloy, Miotti and Dennis, ‘Multinational AGI Consortium (MAGIC)’; Cass-Beggs and others, ‘Framework Convention on Global AI Challenges,’ p.15.} Wasil and others argue that such initiatives may be important in attracting technical talent to institutions overseeing international agreements.\footnote{Akash Wasil and others, ‘Governing dual-use technologies: Case studies of international security agreements \& lessons for AI governance’ (2024) \url{https://doi.org/10.48550/arXiv.2409.02779}, p.10.}

Secondly, distributing AI-related resources and technologies—an example of benefit sharing—could incentivize participation from states not at the forefront of AI development.\footnote{Trager and others, ‘International Governance of Civilian AI,’ p.3. We acknowledge that benefit sharing is not only important as an incentive for states to sign an international agreement. See Claire Dennis and others, ‘Options and Motivations for International AI Benefit Sharing’ (2025) \url{https://www.governance.ai/research-paper/options-and-motivations-for-international-ai-benefit-sharing}} Dennis and others suggest this approach may also appeal to states at the forefront that wish to increase their global market share.\footnote{Ibid., p.10.}  

Thirdly, Hoffman and others note that international treaties have been more effective when they concern trade or finance and contain enforcement mechanisms.\footnote{Steven Hoffman and others, ‘International Treaties have mostly Failed to Produce their Intended Effects’, \textit{PNAS}, 119 (2022) \url{https://www.pnas.org/doi/full/10.1073/pnas.2122854119}} Drawing upon this finding, Trager and others use trade as an incentive in their proposal.\footnote{Trager and others, ‘International Governance of Civilian AI.’}  As discussed in Section 1.3.3, they propose an International Artificial Intelligence Organization (IAIO) to certify states’ compliance with international standards. To be certified, states would have to \emph{enforce} restrictions on trading AI products with uncertified states. This approach is comparable to the EU AI Act, which prohibits non-compliant AI systems from entering the EU internal market. If states at the forefront of AI development, like the US and China, adopted a similar agreement, it could create a powerful incentive for other states to sign.

\section{Recommendations}

\textbf{There are encouraging areas of consensus} across proposals for international agreements, many of which build upon precedents set by existing practices and regulations. 

Developers of frontier models often voluntarily report to AI Safety Institutes (AISIs) and work with private evaluators to assess the risks of their models.\footnote{\textit{Press Release: World Leaders, Top AI Companies Set Out Plan for Safety Testing of Frontier as First Global AI Safety Summit Concludes}, (2023) \url{https://www.gov.uk/government/news/world-leaders-top-ai-companies-set-out-plan-for-safety-testing-of-frontier-as-first-global-ai-safety-summit-concludes}. [accessed Feb 25, 2025]; \textit{U.S. AI Safety Institute Signs Agreements Regarding AI Safety Research, Testing and Evaluation With Anthropic and OpenAI}, (2024) \url{https://www.nist.gov/news-events/news/2024/08/us-ai-safety-institute-signs-agreements-regarding-ai-safety-research} [accessed Feb 25, 2025]; OpenAI, \textit{OpenAI O1 System Card}, p.44.} The joint testing of Anthropic and OpenAI’s models by the UK and US AISIs demonstrates the viability of third-party audits by public bodies and information sharing among international AISIs.\footnote{US AI Safety Institute and UK AI Security Institute, \textit{US AISI and UK AISI Joint Pre-Deployment Test: Anthropic's Claude 3.5 Sonnet}, (2024) \url{https://cdn.prod.website-files.com/663bd486c5e4c81588db7a1d/673b689ec926d8d32e889a8e_UK-US-Testing-Report-Nov-19.pdf}]; US AI Safety Institute and UK AI Security Institute, \textit{US AISI and UK AISI Joint Pre-Deployment Test: OpenAI O1}, (2024) \url{https://cdn.prod.website-files.com/663bd486c5e4c81588db7a1d/6763fac97cd22a9484ac3c37_o1_uk_us_december_publication_final.pdf}.}

\textbf{Practices from high-risk industries are also being adapted to AI development}, as shown by the UK AISI’s collaboration with frontier labs and research organisations to develop safety cases.\footnote{Irving, \textit{Safety Cases at AISI}; Arthur Goemans and others, ‘Safety Case Template for Frontier AI: A Cyber Inability Argument’, (2024) \url{https://doi.org/10.48550/arXiv.2411.08088}. See also Section 1.2.4.} More broadly, the UK AISI’s reporting has set an important precedent for scientific consensus building, while the annual AI summits have created a platform for policy debates.\footnote{Bengio and others, \textit{International AI Safety Report}.}

\textbf{Further research will inform the precise provisions of an international agreement}. We consider compute thresholds a necessary if imperfect regulatory tool. It is a priority to identify the optimal way to calculate them and the level at which they should be set. Additionally, as research on evaluations continues, it will be important to consider how emerging best practices and standards can support the accreditation of private evaluators. Finally, to enable enforcement, hardware-based methods for verifying chip locations and properties of workloads may also need to be developed.\footnote{Reuel, Anka and Ben Bucknall and Anka Reuel, `Open Problems in Technical AI Governance', (2024) \url{https://cdn.governance.ai/Open_Problems_in_Technical_AI_Governance.pdf}, p.26-27.}
This research can be carried out in industry, academia, and by AISIs. Annual reporting on policy-relevant technical matters by AISIs would help inform the provisions and implementation of agreements.

Nevertheless, \textbf{several measures can be implemented more readily}. Given that high-risk industries are commonly subjected to information security and governance audits, it is reasonable to expect the same from AI developers.\footnote{See Section 1.2.2 and Section 1.2.3.} At a minimum, these audits would increase transparency, putting pressure on developers to prioritise safety. Similarly, jurisdictions should also impose Know-Your-Customer requirements and other reporting obligations on cloud compute providers within their borders.\footnote{See Section 1.4.4.}

We advocate for implementing an international agreement, but we recognise that this may require a shift in how the public and policymakers perceive risk. Although many proposals aim to proactively prevent harm, people may only respond \emph{reactively} to evidence of concrete harm. It is therefore crucial to establish incident reporting procedures.\footnote{See for example Ren Bin Lee Dixon and Heather Frase, \textit{An Argument for Hybrid AI Incident Reporting: Lessons Learned from Other Incident Reporting Systems}, (2024) \url{https://cset.georgetown.edu/publication/an-argument-for-hybrid-ai-incident-reporting/} Noam Kolt and others, ‘Responsible Reporting for Frontier AI Development’ (2024), \url{https://doi.org/10.48550/arXiv.2404.02675}; Tommy Shaffer Shane, \textit{AI Incident Reporting: Addressing a Gap in the UK’s Regulation of AI} (2024) \url{https://www.longtermresilience.org/reports/ai-incident-reporting-addressing-a-gap-in-the-uks-regulation-of-ai/}; Merlin Stein and others, ‘The Role of Governments in Increasing Interconnected Post-Deployment Monitoring of AI’ (2024) \url{https://doi.org/10.48550/arXiv.2410.04931}.} Companies should be required, and citizens enabled, to report incidents of harm caused by AI. This could raise awareness of risks and support for measures to address them.

The measures discussed above lay the groundwork for implementing an agreement by advancing technical capabilities for assessment and verification, establishing key practices, and building an evidence base to strengthen political support for addressing risks. Against this backdrop, \textbf{we recommend a conditional AI safety treaty, with the provisions listed below}. The treaty would ideally apply to models developed in the private and public sectors for civilian or military use. To be effective, states parties would need to include the US and China. 

\begin{itemize}

\item \textbf{Establish a compute threshold} above which development should be regulated. 

\begin{itemize}

\item Compute thresholds are the most effective `initial filter' to identify models that \emph{may} present significant risks.\footnote{Heim and Koessler, ‘Training Compute Thresholds,’ p.3.} 
\item AISIs are best placed to specify and revise this threshold and to decide how it is calculated.\footnote{See Section 1.1, footnote 13.}
\item Acknowledging the limitations of compute thresholds, AISIs may also incorporate other metrics in thresholds.
\item Contrary to some proposals,\footnote{Cass-Beggs and others, ‘Framework Convention on Global AI Challenges,’ p.15; Hausenloy, Miotti and Dennis, ‘Multinational AGI Consortium (MAGIC)’; Miotti and Wasil, ‘Taking control,’ p.7; \textit{Treaty on Artificial Intelligence Safety and Cooperation (TAISC)}.} we do not propose exemptions for an international research institute.

\end{itemize}

\item \textbf{Require model audits }(evaluations and red-teaming) for models above the threshold. 

\begin{itemize}

\item These audits should include evaluations conducted by AISIs or other public bodies with grey or white box access to models. Evaluations should occur during development at intervals approved by AISIs.\footnote{See Apollo Research, \textit{Our current policy positions} (2024) \url{https://www.apolloresearch.ai/blog/our-current-policy-positions} [accessed Feb 25, 2025]}
\item AISIs should be convinced that a model does not pose an unacceptable risk, whether due to loss of control, other malfunction risks, or malicious use. If AISIs determine that a model poses unacceptable risk, states parties should ensure its development is paused.
\item AISIs may also require developers to present safety cases, using a range of evidence to prove that risks are kept below quantified levels, as is required in other high-risk industries.\footnote{For example, the U.S. Nuclear Regulatory Commission set a quantitative goal for a Core Damage Frequency of less than 1 × 10\textsuperscript{-4} per year. United States Nuclear Regulatory Commission, ‘Risk Metrics for Operating New Reactors’ (2009) \url{https://www.nrc.gov/docs/ML0909/ML090910608.pdf}.} 

\end{itemize}

\item \textbf{Require security and governance audits for developers of models above the threshold}. Based on practices in high-risk industries, these audits could be performed by accredited private bodies. Governance audits should assess whether a developer has sufficient risk management procedures and a safety culture.\footnote{See Section 1.2.3.}
\item \textbf{Impose reporting requirements and Know-Your-Customer requirements on cloud compute providers}. To the best of their ability, providers should report the amount of compute consumed by their customers to a designated authority.
\item \textbf{Verify implementation via oversight of the compute supply chain}. Verification methods include monitoring transfers of AI-related hardware through customs and financial data analysis, and requirements to report sales and purchases; remote monitoring to identify data centres and energy consumption patterns; hardware components to track chips and their activities (contingent on further research); and on-site routine inspections or challenge inspections. 

\begin{itemize}
\item It may be beneficial for an international institution to conduct inspections to reduce inter-state suspicion and reduce the potential for conflict. See Section 1.4.4 for a discussion of verification methods. 
\item States parties will bear ultimate responsibility for enforcing provisions, including pausing model development where required.
\end{itemize}
\end{itemize}

\section*{Conclusion}

In this paper, we reviewed proposals for international agreements to address malfunction or misuse of advanced AI. We found that compute thresholds are widely supported, though researchers acknowledge their limitations and propose mechanisms to revise them. 

Above these thresholds, researchers advocate supplementing model audits with third-party security and governance audits, drawing on practices from other high-risk industries to establish multiple layers of safety. However, identifying appropriate methods for auditing models remains an open problem, and there is debate as to whether developers should prove the absence of dangerous capabilities or go further by demonstrating affirmative safety. 

Continued research is therefore essential and can be formalised through reporting processes. Standardisation processes can also inform provisions, although, in the interest of time, they may need to occur outside established international standards organisations.

Furthermore, researchers stress the importance of verification methods to ensure compliance, a challenge that may increase if access to compute becomes more diffuse. We identify four broad categories, in increasing order of feasibility: hardware-based verification, inspections, remote sensing, and customs and financial intelligence.\footnote{Wasil and others, ‘Verification methods for international AI agreements’, p.16.} Effective implementation of an agreement may also require incentives, including research and development, benefit-sharing, and trade—of which we believe the latter two are the most promising.

More broadly, researchers' differing views on the feasibility of soft versus hard law reflect different perspectives on political realities. Yet, given the severity of risks and the fact that they cannot be contained within jurisdictional boundaries, we propose a treaty, building upon existing practices in AI development and best practices in other high-risk industries.

\addcontentsline{toc}{section}{Conclusion}
\newpage
\section*{Bibliography}

\begin{hangparas}{.25in}{1}

Alaga, Jide and Jonas Schuett, ‘Coordinated Pausing: An Evaluation-Based Coordination Scheme for Frontier AI Developers’, (2023) \url{https://doi.org/10.48550/arXiv.2310.00374}

Anthropic, \textit{Anthropic's Responsible Scaling Policy}, (2023) \url{https://www-cdn.anthropic.com/files/4zrzovbb/website/1adf000c8f675958c2ee23805d91aaade1cd4613.pdf}

Anthropic, \textit{Challenges in evaluating AI systems} (2023) \url{https://www.anthropic.com/news/evaluating-ai-systems} [accessed Feb 25, 2025]

Apollo Research, \textit{We need a Science of Evals} (2024) \url{https://www.apolloresearch.ai/blog/we-need-a-science-of-evals} [accessed Feb 25, 2025]

Bengio, Yoshua and others, \textit{International AI Safety Report} (2025) \url{https://www.gov.uk/government/publications/international-ai-safety-report-2025} 

Marie Davidsen Buhl and others, ‘Safety cases for frontier AI’ (2024) \url{<https://doi.org/10.48550/arXiv.2410.21572}

Executive Office of the President [Joe Biden], \textit{Safe, Secure, and Trustworthy Development and use of Artificial Intelligence} (2023) \url{https://www.govinfo.gov/content/pkg/FR-2023-11-01/pdf/2023-24283.pdf} 

Casper, Stephen and Carson Ezell, ‘Black-Box Access is Insufficient for Rigorous AI Audits’ (2024) \url{https://doi.org/10.48550/arXiv.2401.14446}

Cass-Beggs, Duncan and others, ‘Framework Convention on Global AI Challenges’, (2024) \url{https://www.cigionline.org/publications/framework-convention-on-global-ai-challenges/} 

Center for AI Safety, \textit{Statement on AI Risk} (2023), \url{https://www.safe.ai/work/statement-on-ai-risk} [accessed Feb 25, 2025]

Council of Europe: Committee of Ministers, \textit{Council of Europe Framework Convention on Artificial Intelligence and Human Rights, Democracy and the Rule of Law}, (2024) \url{https://www.refworld.org/legal/agreements/coeministers/2024/en/148016} 

Dennis, Claire and others, ‘Options and Motivations for International AI Benefit Sharing’ (2025) \url{https://www.governance.ai/research-paper/options-and-motivations-for-international-ai-benefit-sharing}

Dixon, Ren Bin Lee and Heather Frase, \textit{An Argument for Hybrid AI Incident Reporting: Lessons Learned from Other Incident Reporting Systems}, (2024) \url{https://cset.georgetown.edu/publication/an-argument-for-hybrid-ai-incident-reporting/} 

Egan, Janet and Lennart Heim, ‘Oversight for Frontier AI through a Know-Your-Customer Scheme for Compute Providers’, (2023) \url{https://doi.org/10.48550/arXiv.2310.13625}

Elmgren, Karson and Oliver Guest, ‘Chinese AI Safety Institute Counterparts,’ (2024) \url{https://www.iaps.ai/research/china-aisi-counterparts}

‘Regulation (EU) 2024/1689 of the European Parliament and of the Council of 13 June 2024 Laying Down Harmonised Rules on Artificial Intelligence and Amending Regulations,’ \textit{Official Journal} L (2024) \url{https://eur-lex.europa.eu/legal-content/EN/TXT/PDF/?uri=OJ:L_202401689} 

Fajardo, Teresa, ‘Soft Law,’ \textit{Oxford Bibliographies} (2014) \url{https://www.oxfordbibliographies.com/display/document/obo-9780199796953/obo-9780199796953-0040.xml}

Future of Life Institute, \textit{Pause Giant AI Experiments: An Open Letter}, (2023) \url{https://futureoflife.org/open-letter/pause-giant-ai-experiments/} [accessed Feb 25, 2025]

Goemans, Arthur and others, ‘Safety Case Template for Frontier AI: A Cyber Inability Argument’, (2024) \url{https://doi.org/10.48550/arXiv.2411.08088}

Hadfield, Gillian and Jack Clark, ‘Regulatory Markets: The Future of AI Governance’ (2023) \url{https://doi.org/10.48550/arXiv.2304.04914}

Hausenloy, Jason, Andrea Miotti and Claire Dennis, ‘Multinational AGI Consortium (MAGIC): A Proposal for International Coordination on AI’, (2023) \url{https://doi.org/10.48550/arXiv.2310.09217} 

Heim, Lennart, \textit{Inference Compute: GPT-o1 and AI Governance} (2024) \url{https://blog.heim.xyz/inference-compute/} [accessed Feb 25, 2025]

Heim, Lennart and Leonie Koessler, ‘Training Compute Thresholds: Features and Functions in AI Regulation’ (2024) \url{https://doi.org/10.48550/arXiv.2405.10799}

Heim, Lennart and others, ‘Governing Through the Cloud: The Intermediary Role of Compute Providers in AI Regulation,’ (2024) \url{https://www.oxfordmartin.ox.ac.uk/publications/governing-through-the-cloud-the-intermediary-role-of-compute-providers-in-ai-regulation}

\textit{Hiroshima Process International Code of Conduct for Advanced AI Systems}, (2023) \url{https://digital-strategy.ec.europa.eu/en/library/hiroshima-process-international-code-conduct-advanced-ai-systems} 

\textit{Hiroshima Process International Guiding Principles for Advanced AI Systems}, (2023) \url{https://digital-strategy.ec.europa.eu/en/library/hiroshima-process-international-guiding-principles-advanced-ai-system}

Hoffman, Steven and others, ‘International Treaties have mostly Failed to Produce their Intended Effects’, \textit{PNAS} 119 (2022) \url{https://www.pnas.org/doi/full/10.1073/pnas.2122854119}

Hooker, Sara, ‘On the Limitations of Compute Thresholds as a Governance Strategy,’ (2024) \url{https://doi.org/10.48550/arXiv.2407.05694}

International Atomic Energy Agency, \textit{Safety and Security Culture} (2016) \url{https://www.iaea.org/topics/safety-and-security-culture} [accessed Feb 25, 2025]

International Organization for Standardization, \textit{ISO Membership Manual} (2015) \url{https://www.iso.org/publication/PUB100399.html}

Irving, Geoffrey, \textit{Safety Cases at AISI,} (2024) \url{https://www.aisi.gov.uk/work/safety-cases-at-aisi} [accessed Feb 25, 2025]

Klein, Emma and Stewart Patrick, ‘Envisioning a Global Regime Complex to Govern Artificial Intelligence’, (2024) \url{https://carnegieendowment.org/research/2024/03/envisioning-a-global-regime-complex-to-govern-artificial-intelligence?lang=en} 

Kolt, Noam and others, ‘Responsible Reporting for Frontier AI Development’ (2024), \url{https://doi.org/10.48550/arXiv.2404.02675} 

Maas, Matthijs M. and José Jaime Villalobos, ‘International AI Institutions: A Literature Review of Models, Examples, and Proposals,’ \textit{AI Foundations Report} 1 (2023) \url{http://dx.doi.org/10.2139/ssrn.4579773}

Manheim, David, ‘Building a Culture of Safety for AI: Perspectives and Challenges’, (2023) \url{https://papers.ssrn.com/abstract=4491421}

Matz-Lück, Nele, ‘Framework Conventions as Regulatory Tools’, \textit{Goettingen Journal of International Law}, 1 (2009), 439–458 \url{https://papers.ssrn.com/abstract=1535892}

National Institute of Standards and Technology, \textit{A Plan for Global Engagement on AI Standards} (2024) \url{https://doi.org/10.6028/NIST.AI.100-5}

Miotti, Andrea and Akash Wasil, ‘An International Treaty to Implement a Global Compute Cap for Advanced Artificial Intelligence’, (2023) \url{https://doi.org/10.48550/arXiv.2311.10748}

Miotti, Andrea and Akash Wasil, ‘Taking Control: Policies to Address Extinction Risks from Advanced AI’, (2023) \url{https://doi.org/10.48550/arXiv.2310.20563} 

OECD, AI Principles, \url{https://www.oecd.org/en/topics/ai-principles.html} [accessed Feb 25, 2025]

OECD, \textit{Global Partnership on Artificial Intelligence}, \url{https://www.oecd.org/en/about/programmes/global-partnership-on-artificial-intelligence.html} [accessed Feb 25, 2025]

OpenAI, \textit{OpenAI O1 System Card} (2024) \url{https://cdn.openai.com/o1-system-card-20241205.pdf} 

OpenAI, \textit{Preparedness Framework (Beta)}, (2023) \url{https://cdn.openai.com/openai-preparedness-framework-beta.pdf} 

PauseAI, \textit{PauseAI Proposal}, (2024) \url{https://pauseai.info/proposal}

Pouget, Hadrien, ‘What will the role of standards be in AI governance?’(2023) \url{https://www.adalovelaceinstitute.org/blog/role-of-standards-in-ai-governance/} [accessed Feb 25, 2025]

Pouget, Hadrien and Claire Dennis, ‘The Future of International Scientific Assessments of AI’s Risks’, (2024) \url{https://www.oxfordmartin.ox.ac.uk/publications/the-future-of-international-scientific-assessments-of-ais-risks}

Reuel, Anka and others, ‘Position Paper: Technical Research and Talent is Needed for Effective AI Governance,’ \textit{Proceedings of the 41st International Conference on Machine Learning}, (June 11, 2024) \url{https://doi.org/10.48550/arXiv.2406.06987}

Righetti, Luca, Dangerous capability tests should be harder (2024) \url{https://www.planned-obsolescence.org/dangerous-capability-tests-should-be-harder/} [accessed Feb 25, 2025]

Roberts, Huw and others, ‘Global AI Governance: Barriers and Pathways Forward’, \textit{International Affairs}, 100 (2024), 1275–1286 \url{https://doi.org/10.1093/ia/iiae073} 

Sastry, Girish and others, ‘Computing Power and the Governance of Artificial Intelligence’, (2024) \url{https://doi.org/10.48550/arXiv.2402.08797}

Shaffer Shane, Tommy \textit{AI Incident Reporting: Addressing a Gap in the UK’s Regulation of AI} (2024) \url{https://www.longtermresilience.org/reports/ai-incident-reporting-addressing-a-gap-in-the-uks-regulation-of-ai/}

Shevlane, Toby and others, ‘Model Evaluation for Extreme Risks’, (2023) \url{https://doi.org/10.48550/arXiv.2305.15324}

Stein, Merlin and others, ‘Public Vs Private Bodies: Who should Run Advanced AI Evaluations and Audits? A Three-Step Logic Based on Case Studies of High-Risk Industries’, (2024) \url{https://www.oxfordmartin.ox.ac.uk/publications/public-vs-private-bodies-who-should-run-advanced-ai-evaluations-and-audits-a-three-step-logic-based-on-case-studies-of-high-risk-industries}

Stein, Merlin and others, ‘The Role of Governments in Increasing Interconnected Post-Deployment Monitoring of AI’ (2024) \url{https://doi.org/10.48550/arXiv.2410.04931} 

Thurnherr, Lara and others, ‘Who Should Develop Which AI Evaluations,’ (2025) \url{https://www.oxfordmartin.ox.ac.uk/publications/who-should-develop-which-ai-evaluations}

Trager, Robert and others, ‘International Governance of Civilian AI: A Jurisdictional Certification Approach’, (2023) \url{https://doi.org/10.48550/arXiv.2308.15514}

\textit{Treaty on Artificial Intelligence Safety and Cooperation (TAISC)}, \url{https://taisc.org} [accessed Feb 25, 2025]

\textit{Press Release: World Leaders, Top AI Companies Set Out Plan for Safety Testing of Frontier as First Global AI Safety Summit Concludes}, (2023) \url{https://www.gov.uk/government/news/world-leaders-top-ai-companies-set-out-plan-for-safety-testing-of-frontier-as-first-global-ai-safety-summit-concludes} [accessed Feb 25, 2025]

UK AI Security Institute, \textit{Early Lessons from Evaluating Frontier AI Systems}, (2024) \url{https://www.aisi.gov.uk/work/early-lessons-from-evaluating-frontier-ai-systems} [accessed Feb 25, 2025]

UNESCO, \textit{Recommendation on the Ethics of Artificial Intelligence}, (2022) \url{https://unesdoc.unesco.org/ark:/48223/pf0000381137} 

United Nations, \textit{Global Digital Compact}, (2024) \url{https://www.un.org/global-digital-compact/en}

United Nations General Assembly, \textit{United Nations Framework Convention on Climate Change: resolution / adopted by the General Assembly}, (1994) \url{https://unfccc.int/resource/docs/convkp/conveng.pdf}

United States Nuclear Regulatory Commission, \textit{Cybersecurity} (2025) \url{https://www.nrc.gov/security/cybersecurity.html} [accessed Feb 25, 2025]

United States Nuclear Regulatory Commission, ‘Risk Metrics for Operating New Reactors’ (2009) \url{https://www.nrc.gov/docs/ML0909/ML090910608.pdf} 

US AI Safety Institute and UK AI Security Institute, \textit{US AISI and UK AISI Joint Pre-Deployment Test: Anthropic's Claude 3.5 Sonnet}, (2024) \url{https://cdn.prod.website-files.com/663bd486c5e4c81588db7a1d/673b689ec926d8d32e889a8e_UK-US-Testing-Report-Nov-19.pdf}

US AI Safety Institute and UK AI Security Institute, \textit{US AISI and UK AISI Joint Pre-Deployment Test: OpenAI O1}, (2024) \url{https://cdn.prod.website-files.com/663bd486c5e4c81588db7a1d/6763fac97cd22a9484ac3c37_o1_uk_us_december_publication_final.pdf}

Villalobos, José Jaime and Matthijs M. Maas, ‘Beyond a Piecemeal Approach: Prospects for a Framework Convention on AI’, \textit{SSRN}. Forthcoming in \textit{The Oxford Handbook on the Foundations and Regulation of Generative AI}, ed. by P Hacker, A Engel, S Hammer and B Mittelstadt, (Oxford: Oxford University Press, 2024), \url{https://papers.ssrn.com/abstract=5020616} 

Wasil, Akash and others, ‘Affirmative Safety: An Approach to Risk Management for High-Risk AI’, (2024) \url{https://doi.org/10.48550/arXiv.2406.15371} 

Wasil, Akash and others, ‘Governing dual-use technologies: Case studies of international security agreements \& lessons for AI governance’ (2024) \url{https://doi.org/10.48550/arXiv.2409.02779}

Wasil, Akash and others, ‘Verification methods for international AI agreements’, (2023) \url{https://doi.org/10.48550/arXiv.2408.16074}

Wiener, \textit{Scott Safe and Secure Innovation for Frontier Artificial Intelligence Models Act}, (2024) \url{https://leginfo.legislature.ca.gov/faces/billTextClient.xhtml?bill_id=202320240SB1047}

Ziosi, Marta and others, ‘AISIs’ Roles in Domestic and International Governance’ (2024) \url{https://www.oxfordmartin.ox.ac.uk/publications/aisis-roles-in-domestic-and-international-governance}

\addcontentsline{toc}{section}{Bibliography}
\end{hangparas}

\end{document}